\documentclass{article}
\usepackage{spconf,amsmath,graphicx}
\usepackage{amsfonts}
\usepackage{bbm}

\title{GLOBAL-CONTEXT AWARE GENERATIVE PROTEIN DESIGN}
\name{Cheng Tan$^{1,2,3\dag}$ \qquad Zhangyang Gao$^{1,2,3\dag}$ \qquad Jun Xia$^{1,2,3}$ \qquad Bozhen Hu$^{1,2,3}$ \qquad Stan Z. Li$^{1,2 *}$\thanks{*Corresponding author}\thanks{\dag Equal contribution}}
  
\address{
    $^{1}$ Zhejiang University.
    $^{2}$ AI Lab, School of Engineering, Westlake University. \\
    $^{3}$ Institute of Advanced Technology, Westlake Institute for Advanced Study 
}
\begin{document}
%
\maketitle
\begin{abstract}
The linear sequence of amino acids determines protein structure and function. Protein design, known as the inverse of protein structure prediction, aims to obtain a novel protein sequence that will fold into the defined structure. Recent works on computational protein design have studied designing sequences for the desired backbone structure with local positional information and achieved competitive performance. However, similar local environments in different backbone structures may result in different amino acids, which indicates the global context of protein structure matters. Thus, we propose the \textbf{G}lobal-\textbf{C}ontext \textbf{A}ware generative de novo protein design method (GCA), consisting of \emph{local} modules and \emph{global} modules. While \emph{local} modules focus on relationships between neighbor amino acids, \emph{global} modules explicitly capture non-local contexts. Experimental results demonstrate that the proposed GCA method achieves state-of-the-art performance on structure-based protein design. Our code and pretrained model have been released on Github\footnote{github.com/chengtan9907/gca-generative-protein-design}.
\end{abstract}
\section{Introduction}
\label{sec:intro}

Computational protein design, which aims to invent protein molecules with desired structures and functions automatically, has a wide range of applications in therapeutics and pharmacology~\cite{brunette2015exploring,huang2016coming,langan2019novo}. Recent years have witnessed remarkable advancements in this field with increased computation power, in which many of them are led by deep learning techniques~\cite{GAO2020100142, casp, yang2019machine}. While classical protein design approaches depend on composite energy functions of protein physics and sampling algorithms for exploring both sequence and structure spaces, data-driven approaches take advantage of deep neural networks to generate protein sequences with less complex prior knowledge. 

Designing a protein sequence for a given structure remains challenging, as the difficulty in mapping the 3D space of structures to the vast-size sequence space. Current data-driven protein design methods~\cite{nips2019_ingraham, jing2021learning, protein_sovler, fold2seq} agree on the assumption based on biology and physics prior knowledge that, for each amino acid, its neighborhoods have the most immediate and vital effects on itself. The majority of such methods represent protein structures as graphs with hand-crafted features and aggregate local messages in hidden layers. The computational protein design process is formulated to learn valuable features from 3D structures with the local message passing mechanism. However, \textit{the similar local environment in different proteins may correspond to different amino acids}. Local neighbors do matter~\cite{ribeiro2020modeling}, but it is not enough to obtain high-quality protein sequences.

\begin{figure}[ht]
    \centering
    \includegraphics[width=0.42\textwidth]{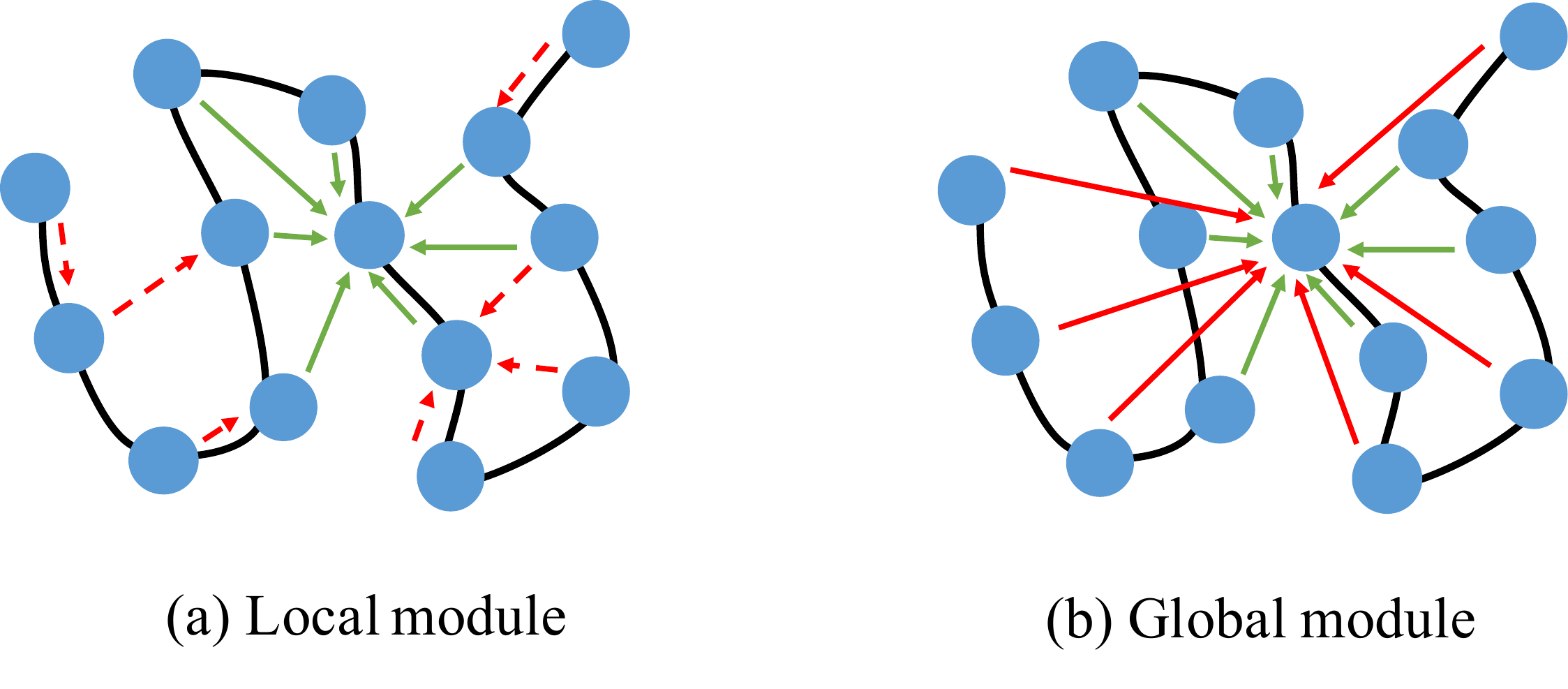}
    \caption{The comparison between the local module and global module. Information flows from adjacent nodes, and distant nodes are denoted as green arrows and red arrows, respectively. The red dashed arrows indicate the implicit information flow from distant nodes.}
    \label{fig:local_global_module}
\end{figure}

To fully explore the non-local information, we propose the \textbf{G}lobal-\textbf{C}ontext \textbf{A}ware generative de novo protein design method (GCA) with both local and global modules. While local modules are built upon graph attention networks that aggregate local messages gained from neighbors with different weights, global modules extend local graph attention to global self-attention neural networks in the form of Transformer~\cite{vaswani2017attention} architectures. As shown in Fig.~\ref{fig:local_global_module}, the local module focuses on adjacent structure information though distant nodes can deliver information implicitly; the global module explicitly gathers information from distant nodes in a self-attention mechanism. By composing multiple blocks of local modules and global modules, our proposed GCA can capture high-order dependencies between protein sequences and protein structures in both neighbor-level and overall-level.

This paper is organized as follows: we introduce the details of our proposed GCA generative de novo protein design method in section 2. Experimental results are reported in section 3, and we conclude in section 4.

\section{Proposed method}
\label{sec:method}

\subsection{Preliminaries}

\begin{figure}[ht]
    \centering
    \includegraphics[width=0.42\textwidth]{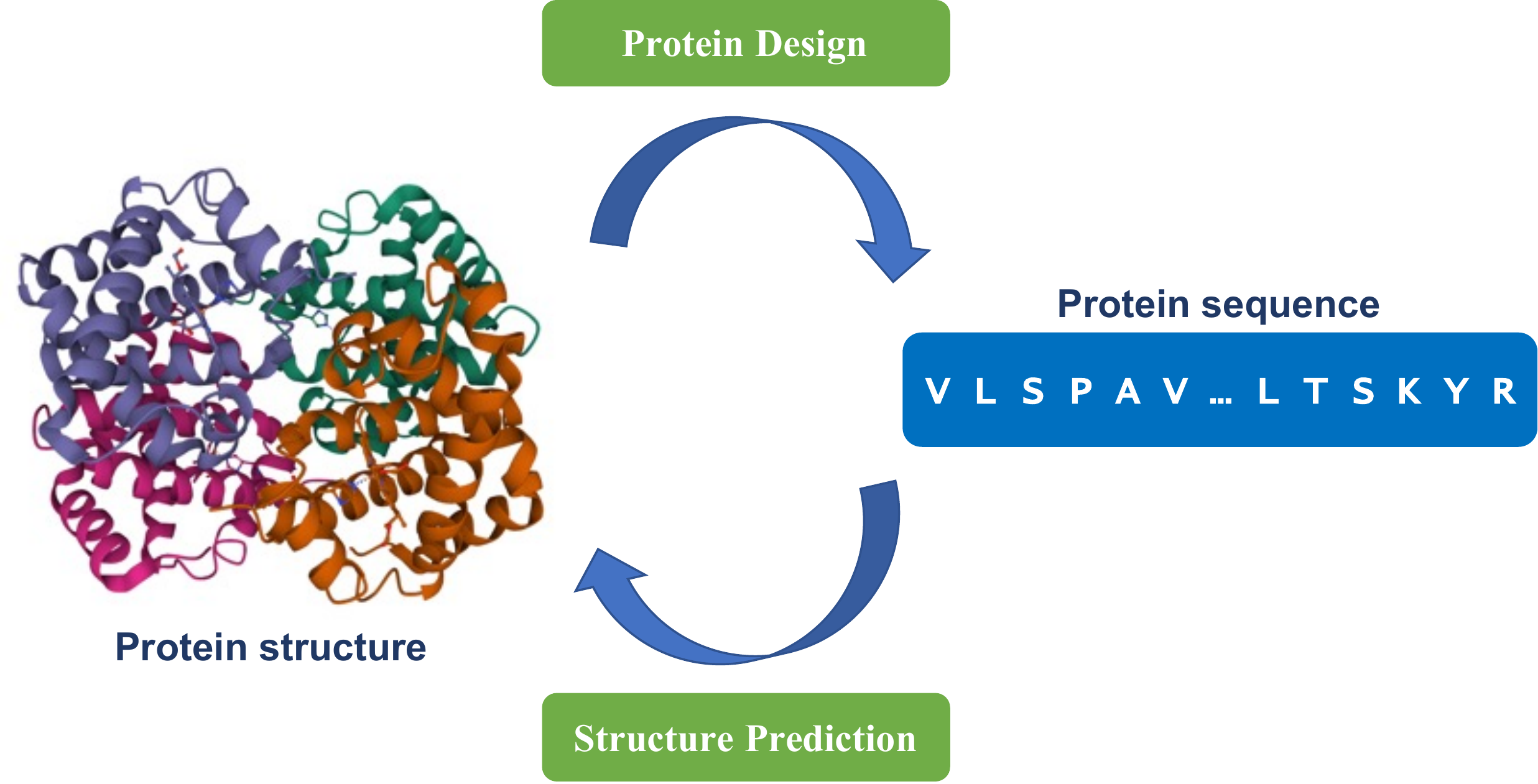}
    \caption{The comparison of two important tasks in protein modeling: structure prediction and protein design. The 3D visualization is created by Mol* Viewer~\cite{sehnal2021mol}.}
    \label{fig:task}
\end{figure}

Protein primary structure is the linear sequence of amino acids, typically notated as a string of letters. A protein sequence $\mathcal{S}^N = \{(a)^N | a \in \{A, R, N, ..V\}\}$ has $N$ amino acids while each of them is represented by a letter of twenty possible letters such as A, R, N, D, C, Q, E, G, H, I, L, K, M, F, P, S, T, W, Y, and V. A protein sequence will fold into a protein tertiary structure $\mathcal{X}^N = \{\mathbf{x}_i^{\omega} \in \mathbb{R}^3: 1 \leq i \leq N, \omega \in \{\mathrm{C}\alpha, \mathrm{C}, \mathrm{N}, \mathrm{O}\}\}$, where $N$ is the number of amino acids and $\omega$ indicates the chain in protein. As shown in Fig.~\ref{fig:task}, protein design task predicts the protein sequence of a given protein structure, while structure prediction task is the opposite.

\subsection{Represent protein as a graph}

The structure of a protein is represented as a graph $\mathcal{G} = (\mathcal{V}, \mathcal{E})$ where node feature $\mathbf{v}_i \in \mathcal{V}$ corresponds to an amino acid while edge feature $\mathcal{E} = \{ \mathbf{e}_{ij} \}_{j \in \mathcal{N}_i}$ suggests the rotation-invariant and translation-invariant relationships between each pair of nodes $\mathbf{v}_i$ and $\mathbf{v}_j$. In particular, $\mathcal{N}_i$ denotes the $K$-nearest neighbors of node $i$ calculated by Euclidean distances of the backbone. 

For node features, we construct three dihedral angles $\{\phi_i, \psi_i, \omega_i \}$ of the protein backbone from $\mathrm{C}_{i-1}, \mathrm{N}_i, \mathrm{C}\alpha_i, \mathrm{C}_i$, and $\mathrm{N}_{i+1}$. Then these dihedral angels are embedded on the 3-torus as $\mathbf{v}_i = \{\sin, \cos\} \times \{\phi_i, \psi_i, \omega_i \}$.

\begin{figure}
    \centering
\includegraphics[width=0.35\textwidth]{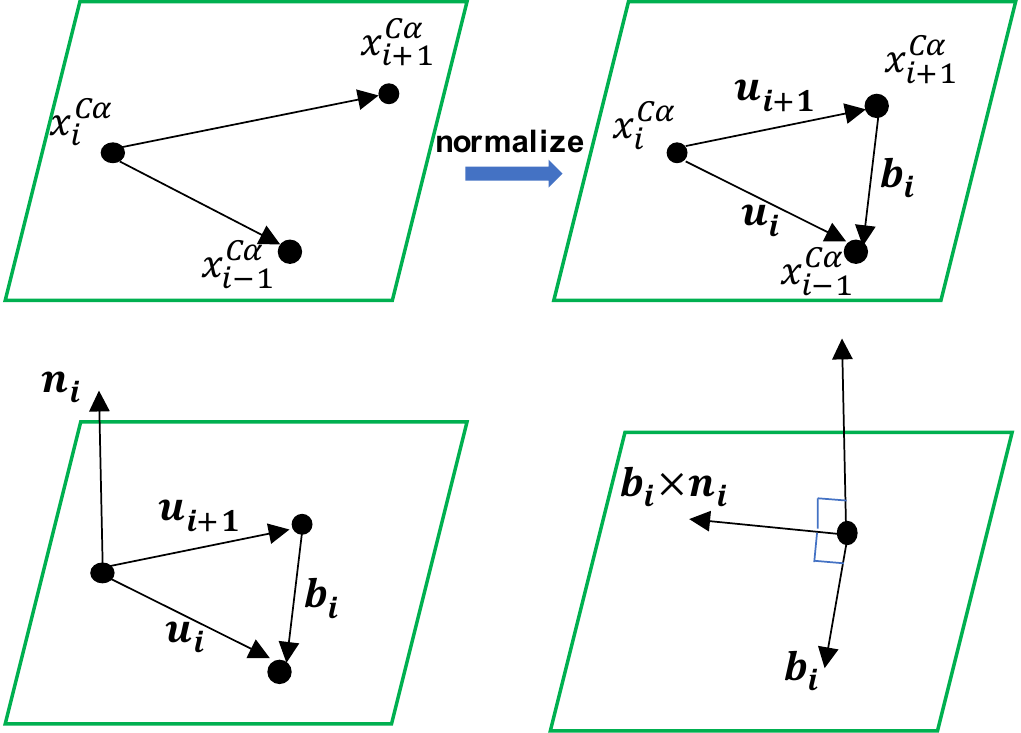}
    \caption{A general view of how the local coordinate system is built. The final coordinate satisfies $\mathbf{b}_i \perp \mathbf{n}_i, \mathbf{b}_i \times \mathbf{n}_i \perp \mathbf{n}_i, \mathbf{b}_i \times \mathbf{n}_i \perp \mathbf{b}_i$.}
    \label{fig:local_coordinate}
\end{figure}

For edge features, we focus on describing relative spatial relationships between amino acids that satisfy rotation-invariant and translation-invariant properties. To simplify the computation, we only consider the position $\mathbf{x}_i^{\mathrm{C}\alpha}$ of the alpha carbon $\mathrm{C}\alpha$ as it's the central carbon atom in each amino acid. The distance $\|\mathbf{x}_j^{C\alpha}-\mathbf{x}_i^{C\alpha}\|_2, \forall i \neq j$ is encoded by Gaussian radial basis functions $\mathbf{r}(\cdot)$. Then, as shown in Fig.~\ref{fig:local_coordinate}, the direction is encoded by $\mathbf{O}_i^T \frac{\mathbf{x}_j^{C\alpha} - \mathbf{x}_i^{C\alpha}}{\|\mathbf{x}_j^{C\alpha} - \mathbf{x}_i^{C\alpha}\|}$ while $\mathbf{O}_i = [\mathbf{b}_i \quad \mathbf{n}_i \quad \mathbf{b}_i \times \mathbf{n}_i]$ defines a local coordinate system for each amino acid by:

\begin{equation}
    \mathbf{u}_i = \frac{\mathbf{x}_i^{C\alpha} - \mathbf{x}_{i-1}^{C\alpha}}{\|\mathbf{x}_i^{C\alpha} - \mathbf{x}_{i-1}^{C\alpha}\|},
    \mathbf{b}_i = \frac{\mathbf{u}_i - \mathbf{u}_{i+1}}{\|\mathbf{u}_i - \mathbf{u}_{i+1}\|},
    \mathbf{n}_i = \frac{\mathbf{u}_i \times \mathbf{u}_{i+1}}{\|\mathbf{u}_i \times \mathbf{u}_{i+1}\|}.
\end{equation}
The orientation is encoded by the common-used quaternion representation of rotation matrix $\mathbf{q}(\mathbf{O}_i^T \mathbf{O}_j)$. Thus, the edge feature $\mathbf{e}_{ij}$ is the concatenation of the distance, direction and orientation encodings as:

\begin{equation}
    \mathbf{e}_{ij} = \Big(\mathbf{r}(\|\mathbf{x}_j^{C\alpha}-\mathbf{x}_i^{C\alpha}\|_2), \mathbf{O}_i^T \frac{\mathbf{x}_j^{C\alpha} - \mathbf{x}_i^{C\alpha}}{\|\mathbf{x}_j^{C\alpha} - \mathbf{x}_i^{C\alpha}\|}, \mathbf{q}(\mathbf{O}_i^T \mathbf{O}_j) \Big).
\end{equation}

\subsection{Network architecture}

\subsubsection{Local module}
The local module is a graph neural network (GNN) that aggregates both node embeddings and local edge embeddings and updates the node embedding for further sequence generations. Considering a $L$-layer GNN, the key operations \emph{aggregating} and \emph{updating} can be formulated as follows:

\begin{equation}
    \mathbf{h}_{\mathcal{N}_i}^{(l)} = \mathit{aggregating}^{(l)}(\{ \big(\mathbf{h}_i^{(l-1)}, \mathbf{h}_j^{(l-1)}, \mathbf{h}_{\mathbf{e}_{ij}} \big): j \in \mathcal{N}_i\}),
\end{equation}

\begin{equation}
    \mathbf{h}_i^{(l)} = \mathit{updating}(\mathbf{h}_i^{(l-1)}, \mathbf{h}_{\mathcal{N}_i}^{(l)}),
\end{equation}
where $\mathbf{h}_i^{(l)} \in \mathbb{R}^{D}$ denotes the embedding of node $i$ on the $l$-th layer, $\mathbf{h}_{\mathcal{N}_i}^{(l)} \in \mathbb{R}^{K \times D}$ denotes the local edge embedding of node $i$'s neighbors on the $l$-th layer, $K$ is the number of local neighbors, and $D$ is the dimensions of the embedding. In particular, $\mathbf{h}_i^{(0)} \in \mathbb{R}^D$ is the embedding of $\mathbf{v}_i$, and $\mathbf{h}_{\mathbf{e}_{ij}} \in \mathbb{R}^{D}$ is the embedding of the edge feature $\mathbf{e}_{ij}$. The local edge information flows into node embeddings at each layer, while distant edge information flows through high-level layers.

In order to capture the relationships in local neighborhoods, we generalize graph attention scheme that take advantage of attention coefficients $\alpha \in \mathbb{R}^{N \times K}$ as strong relational inductive bias. Specifically, the attention coefficients are calculated as follows:

\begin{equation}
    \alpha_{ij} = \frac{\exp(c_{ij})}{\sum_{k \in \mathcal{N}(i)} \exp(c_{ik})}, \forall j \in \mathcal{N}_i,
\end{equation}
where $c_{ij}$ is expressed as:

\begin{equation}
    c_{ij} = \sigma\Big(a^T \big[\mathbf{W} \mathbf{h}_i^{(l-1)} \; || \; \mathbf{W} \mathbf{h}_j^{(l-1)} \; || \; \mathbf{W}  \mathbf{h}_{\mathbf{e}_{ij}} \big] \Big), \forall j \in \mathcal{N}_i,
\end{equation}
and $\mathbf{W} \in \mathbb{R}^{D \times D}$, $a \in \mathbb{R}^{3D}$ are learnable parameters, $\sigma$ is the activation function, $||$ is the concatenation operation.

Thus, the \emph{aggregating} operation is adopted as:

\begin{equation}
    \mathbf{h}_{\mathcal{N}_i}^{(l)} = \sum_{j \in \mathcal{N}_i} \alpha_{ij} \mathbf{W}_r \big[\mathbf{h}_i^{(l-1)} \; || \; \mathbf{h}_j^{(l-1)} \; || \; \mathbf{h}_{\mathbf{e}_{ij}} \big],
\end{equation}
where $\mathbf{W}_r \in \mathbb{R}^{D \times 3D}$ encodes the relation between $i$ and $j$. The \emph{updating} operation is simply renovating hidden layers by their local neighbors: $\mathbf{h}^{(l)}_i = \mathbf{h}_{\mathcal{N}_i}^{(l)}$.

\subsubsection{Global module}
The global module is the fully self-attention network that generalizes Transformer~\cite{vaswani2017attention} to protein graph. Specifically, the attention coefficients are calculated as follows:

\begin{equation}
    \alpha_{ij} = \frac{\exp(c_{ij})}{\sum_{k \in \mathcal{V}} \exp(c_{ik})},
\end{equation}
where $c_{ij}$ is expressed as:

\begin{equation}
    c_{ij} = \frac{1}{\sqrt{d}} \Big(\mathbf{W}_q \mathbf{h}_i^{(l-1)} \Big)^T \Big(\mathbf{W}_k \big[\mathbf{h}_i^{(l-1)} \; || \; \mathbf{h}_j^{(l-1)} \; || \; \mathbf{h}_{\mathbf{e}_{ij}} \big] \Big),
\end{equation}
where $\mathbf{W}_q \in \mathbb{R}^{D \times D}, \mathbf{W}_k \in \mathbb{R}^{D \times 3D}$ are parameter matrices for the query and key, and $d$ is a scale factor.

Then, the \emph{aggregating} operation is formulated as:
\begin{equation}
    \mathbf{h}_{\mathcal{N}_i}^{(l)} = \sum_{j \in \mathcal{V}} \alpha_{ij} \mathbf{W}_r \mathbf{h}_j^{(l-1)},
\end{equation}
the \emph{updating} operation is defined by employing layer normalization (LayerNorm), dropout (DropOut) and fully connected networks (FFN):
\begin{equation}
    \mathbf{h}^{(l)}_i = \mathrm{LayerNorm}(\mathbf{h}_{\mathcal{N}_i}^{(l)} + \mathrm{DropOut}(\mathrm{FFN}(\mathbf{h}_{\mathcal{N}_i}^{(l)}))).
\end{equation}

\begin{figure*}[ht]
    \centering
    \includegraphics[width=0.99\textwidth]{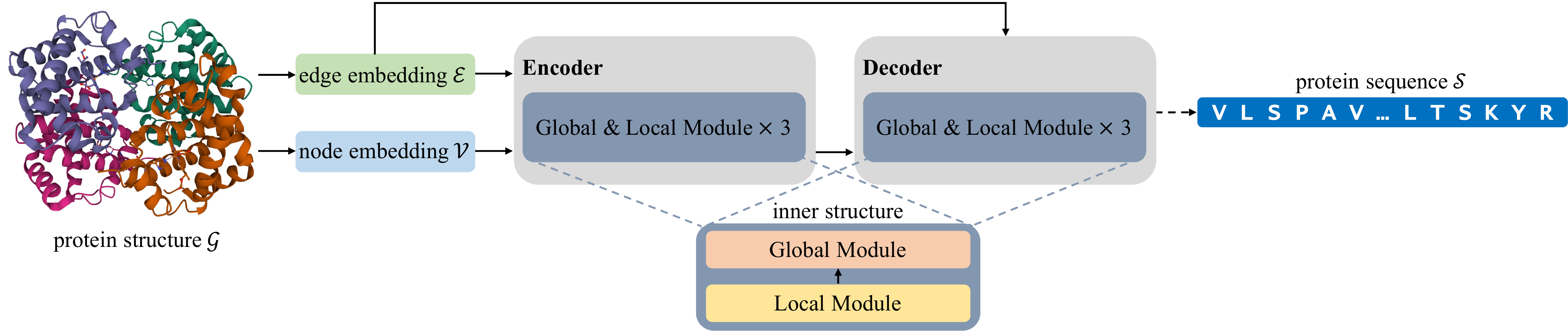}
    \caption{The architecture of our proposed method.}
    \label{fig:architecture}
\end{figure*}

The overall architecture with stacked local modules and global modules is shown in Fig.~\ref{fig:architecture}.

\section{Experiments}
\label{sec:experiments}

\subsection{Experimental settings}


\subsubsection{Dataset}
We use the CATH 4.2 dataset collected by~\cite{nips2019_ingraham} to evaluate the ability of our method to generalize across different protein folds. This dataset obtains full chains up to length 500, and structures have been partitioned with 40\% non-redundancy by their CATH (\textbf{C}lass, \textbf{A}rchitecture, \textbf{T}opology, \textbf{H}omologous) for all domains. As the evaluation set and the test set have minor similarities to the training set, we consider this dataset is approximate to the real-world scenarios that require the design of novel structures.  With no CAT overlap between sets, there are 18024 chains in the training set, 608 chains in the validation set, and 1120 chains in the test set, respectively. Two subsets of the entire test set are evaluated simultaneously: a 'Short' subset containing chains up to length 100 and a 'Single chain' subset for comparing with baselines that only use the single chain. We also consider a smaller dataset TS50, which is the standard benchmark introduced by \cite{li2014direct}. Though the model is still trained on the CATH 4.2 dataset, we filter the training and validation sets to ensure there is no overlap with TS50.

\subsubsection{Measurement}

\noindent{\textbf{Perplexity}}
Following~\cite{nips2019_ingraham,madani2020progen}, we define the perplexity that evaluates the predicted protein sequences from natural language perspective:

\begin{equation}
    \mathrm{PERP}(\mathcal{S}^N, \mathcal{X}^N) = \exp \Big( -\frac{1}{N} \sum_{i=1}^N \mathcal{S}^N_i \log p \big(\mathcal{S}^N_i \; | \; \mathcal{X}^N_i \big) \Big),
\end{equation}
where $(\mathcal{S}^N, \mathcal{X}^N)$ is the sequence-structure pair of a protein with $N$ amino acids. $\mathcal{S}_i^N, \mathcal{X}_i^N$ denote the $i$-th amino acid in sequence and structure respectively. $p(\mathcal{S}_{i}^N | \mathcal{X}_i^N)$ is the output probability from the model. 

\noindent{\textbf{Recovery}}
To evaluate the predicting accuracy of the protein sequence at per-residue level, we consider the recovery:

\begin{small}
\begin{equation}
    \mathrm{REC(\mathcal{D})} = \frac{1}{|\mathcal{D}|} \sum_{(\mathcal{X}^N, \mathcal{S}^N)\in \mathcal{D}} \frac{1}{N} \sum_{i=1}^N \mathbbm{1}[S^N_i = \arg\max p(S^N_i \; | \; \mathcal{X}^N_i)],
\end{equation}
\end{small}
where $\mathcal{D}$ denotes the whole dataset.



\subsubsection{Model architecture and optimization}

In all experiments, GCA model is built by three blocks of both local modules and global modules for the encoder and decoder with the hidden dimension of 128. The Adam optimizer with learning rate of $1e^{-3}$ is employed. Models are trained for 100 epochs while the sequence of each batch contains up to 2,500 characters.

\subsection{Experimental results}

We first present the median of PERP in Table~\ref{tab:perp}. While the structure-free language model LSTMs produce confusing protein sequences, structure-based models obtain less-perplex protein sequences, indicating the importance of structural features. GCA outperforms other structure-based models as global contexts of protein structures are taken into account.

\begin{table}[ht]
\centering
\setlength{\tabcolsep}{2.2mm}{
\begin{tabular}{lrrr}
\hline
Methods     & Short  & Single chain  & All    \\
\hline
\textbf{Language models} & & & \\
LSTM ($h=128$)        & 16.06  & 16.38  & 17.13  \\
LSTM ($h=256$)        & 16.08  & 16.37  & 17.12  \\
LSTM ($h=512$)        & 15.98  & 16.38  & 17.13  \\
SPIN2       & 12.11  & 12.61  & -      \\
\textbf{Structure-based models} & & & \\
StructTrans & 8.56   & 8.97   & 7.14   \\
StructGNN   & 8.40   & 8.84   & 6.69   \\
GCA  &  \textbf{7.68} & \textbf{8.09} & \textbf{6.44} \\
\hline
\end{tabular}}
\caption{Performance of different methods on CATH 4.2 dataset assessed by PERP (lower is better).}
\label{tab:perp}
\end{table}

Though PERP matters from the perspective of natural language, REC that evaluates the ability of models in inferring sequences given determined structures is also crucial. We compare GCA with other structure-based models in Table~\ref{tab:rec}.
\begin{table}[ht]
\centering
\setlength{\tabcolsep}{4.5mm}{
\begin{tabular}{lrrr}
\hline
Methods    & Short       & Single chain     & All    \\
\hline
StructTrans & 31.59 & 30.35 & 33.90 \\
StructGNN   & 30.90 & 30.85 & 35.25 \\
GCA  & \textbf{33.25} & \textbf{33.04} & \textbf{36.11} \\
\hline
\end{tabular}}
\caption{Performance of different methods on CATH 4.2 dataset assessed by REC (higher is better).}
\label{tab:rec}
\end{table}

GCA obtains the highest REC on all three sets among these structure-based methods. Moreover, the recovery of StructGNN and StructTrans drops significantly in 'Short' and 'Single chain' sets, which suggests they are overfitting on long sequences and multiple chains, while GCA performs consistently well on them. As few structural features can be explored in short sequnce and single chain, the prediction is relatively difficult. However, the global information in GCA makes up for the deficiency of structural features of short chains, making performance significantly improved.

To compare with other methods, we conduct experiments on the standard TS50 dataset and show the results in Table~\ref{tab:ts50}. The methods for comparison include the CNN-based ProDCoNN~\cite{zhang2020prodconn}, the distance-map-based SPROF~\cite{chen2019improve}, the graph-based GVP~\cite{jing2021learning} the sequential representation method SPIN~\cite{li2014direct} and SPIN2~\cite{o2018spin2}, the constraint satisfaction method ProteinSolver~\cite{protein_sovler}, and the popular method Rosetta. GCA achieves remarkable performance and outperforms other methods by a large margin.

\begin{table}[ht]
    \centering
    \setlength{\tabcolsep}{14mm}{
    \begin{tabular}{lrrr}
    \hline
    Methods    & REC \\
    \hline
    Rosetta       & 30.0   \\
    SPIN          & 30.3   \\
    ProteinSolver & 30.8   \\
    SPIN2         & 33.6   \\
    StructTrans   & 36.1   \\
    StructGNN     & 38.0   \\
    SPROF         & 39.2   \\
    ProDCoNN      & 40.7   \\
    GVP           & 44.1   \\
    GCA           & \textbf{47.0} \\
    \hline
    \end{tabular}}
    \caption{Performance of different methods on TS50 dataset assessed by REC (higher is better).}
    \label{tab:ts50}
\end{table}
\vspace{-8mm}

\section{Conclusion}
\label{sec:conclusion}
We introduce the consideration of global information and propose the global-context aware generative de novo protein design method, consisting of local modules and global modules. The local module propagates neighborhood messages across layers, and the global module emphasizes long-term dependencies. Experimental results show that GCA outperforms state-of-the-art methods on benchmark datasets. In 'Short' and 'Single chain' sets, the global-context aware mechanism significantly improves the performance, indicating the potentials to promote structure-based protein design.



\bibliographystyle{IEEEbib}
\bibliography{refs}

\end{document}